\documentclass[aps,superscriptaddress,pre,reprint,amsmath,floatfix]{revtex4-1}
\usepackage{graphicx}
\usepackage{array}
\usepackage{booktabs}
\usepackage[colorlinks, linkcolor=blue]{hyperref} 
\usepackage{bm}
\usepackage{color}
\usepackage{subfigure}
\usepackage{soul}
\usepackage{amsmath}
\usepackage{amsfonts}
\usepackage{amssymb}
\usepackage{lipsum}

\definecolor{darkBlue}{rgb}{0,0,0.6}
\definecolor{darkRed}{rgb}{0.5,0,0}
\definecolor{darkGreen}{rgb}{0,0.5,0}

\begin{document}
\title{Hard-needle elastomer in one spatial dimension}
\author{Danilo B. Liarte}
\email{danilo.liarte@ictp-saifr.org}
\affiliation{ICTP South American Institute for Fundamental Research, S\~ao Paulo, SP, Brazil}
\affiliation{Institute of Theoretical Physics, S\~ao Paulo State University, S\~ao Paulo, SP, Brazil}
\author{Alberto Petri}
\affiliation{CNR–Istituto dei Sistemi Complessi, Dipartimento di Fisica, Universit\`a Sapienza, 00185 Roma, Italy}
\affiliation{ICTP South American Institute for Fundamental Research, S\~ao Paulo, SP, Brazil}
\author{Silvio R. Salinas}
\affiliation{Institute of Physics, University of S\~ao Paulo, S\~ao Paulo, SP, Brazil}
\date{\today}

\begin{abstract}
We perform exact Statistical Mechanics calculations for a system of elongated objects (hard needles) that are restricted to translate along a line and rotate within a plane, and that interact via both excluded-volume steric repulsion and harmonic elastic forces between neighbors.
This system represents a one-dimensional model of a liquid crystal elastomer, and has a zero-tension critical point that we describe using the transfer-matrix method.
In the absence of elastic interactions, we build on previous results by Kantor and Kardar, and find that the nematic order parameter $Q$ decays linearly with tension $\sigma$.
In the presence of elastic interactions, the system exhibits a standard universal scaling form, with $Q / \vert\sigma\vert$ being a function of the rescaled elastic energy constant $k / \vert\sigma\vert^\Delta$, where $\Delta$ is a critical exponent equal to $2$ for this model.
At zero tension, simple scaling arguments lead to the asymptotic behavior $Q \sim k^{1/\Delta}$, which does not depend on the equilibrium distance of the springs in this model. 
\end{abstract}

\maketitle

\section{Introduction}

One-dimensional models have often been invoked to illustrate diverse subtleties regarding statistical features of systems in physical dimensions~\cite{Mattis1992}.
Even though strong fluctuations usually prevent the emergence of long-range ordered phases at finite temperature~\cite{Salinas2001}, these models have the major advantage that they can often be solved exactly, and they are amenable to approaches as the renormalization group~\cite{Cardy1996}, leading to the description of far-reaching universal scaling features.
Unsurprisingly, these models provide valuable tools to describe the rich criticality of systems such as strongly-correlated quantum systems, where theoretical progress in dimension higher than one is hindered by formidable analytical and numerical challenges.

A few years ago, Kantor and Kardar have used analytical and numerical calculations to describe the statistical properties of a one-dimension gas of hard anisotropic bodies (ellipsoids, needles, rectangles, etc.) with excluded volume interactions~\cite{KantorKar2009a,KantorKar2009b} (see also Ref.~\cite{LebowitzTal1987}).
The glassy dynamics of a class of similar models was investigated by Arenzon and colleagues~\cite{ArenzonDic2011}, and a two-dimensional gas of hard needles was simulated by Vink~\cite{Vink2009}.
Also, a more recent analysis for a lattice model of hard rotors was carried out by Dhar and colleagues~\cite{SaryalDha2022,KlamserDha2022}.
In this manuscript, we revisit the work by Kantor and Kardar, with the addition of elastic degrees of freedom. 
For Ising systems, the incorporation of compressibility through the addition of harmonic elastic interactions can change the critical behavior and give rise to multicritical points~\cite{Salinas1973,LiarteYok2009}.
In turn, here we make contact with liquid crystal elastomers~\cite{WarnerTer2003}, where the coupling between elastic and orientational degrees of freedom leads to a highly versatile material combining the properties of both rubber and liquid crystals.

Liquid crystal elastomers have attracted  much attention since de Gennes' pioneering paper~\cite{Gennes1969}.
Previous theoretical approaches to describe the intriguing properties of these anisotropic polymer networks include Warner and Terentjev {\it neo-classical theory of elasticity}~\cite{WarnerTer2003}, the lattice version of the Warner-Terentjev model proposed by Selinger and Ratna~\cite{SelingerRat2004} and the minimal models considered by Ye and Lubensky~\cite{YeLub2009}.
In previous publications, some of us have used a mean-field (infinite-range) lattice model (akin to Selinger and Ratna's model) to describe the nematic-isotropic, the poli to mono-domain as well as soft transitions~\cite{LiarteYok2011,Liarte2013,LiarteSal2014,PetriSal2018}.

\begin{figure}[!ht]
\includegraphics[width=\linewidth]{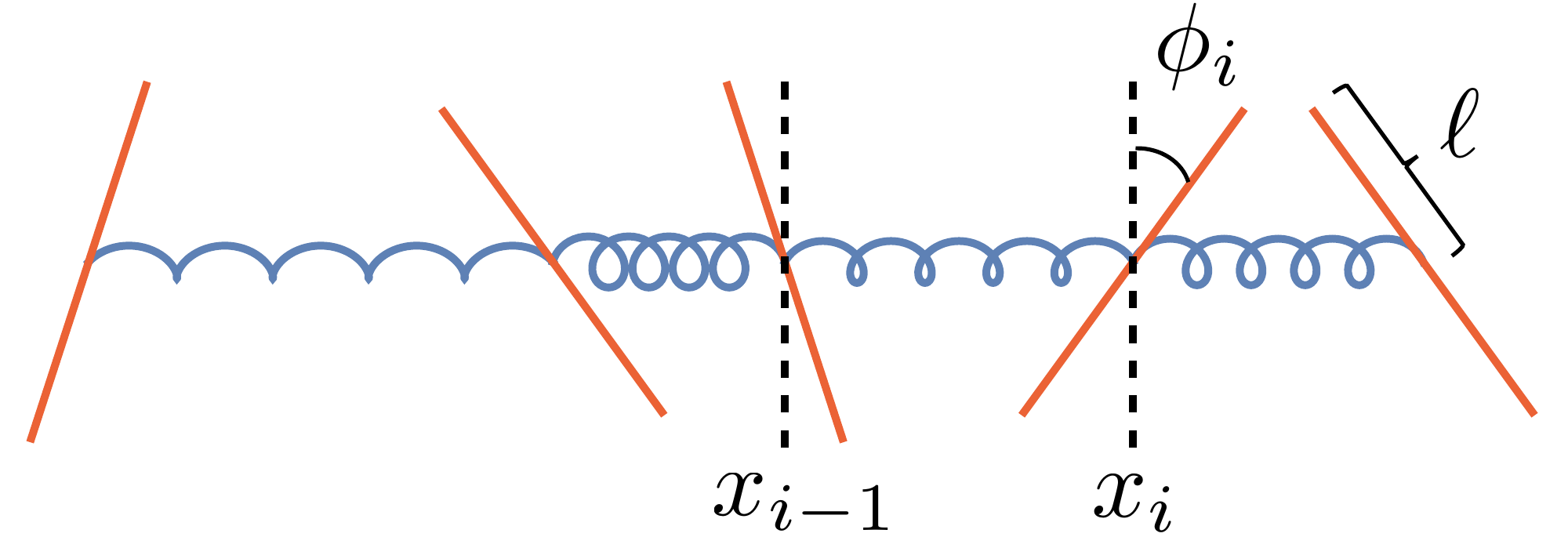}
\caption{Sketch of a set of hard needles with steric repulsions and harmonic elastic interactions.
\label{fig:needles}}
\end{figure}

A chain of hard needles is illustrated in Figure~\ref{fig:needles}.
We consider a system of $N$ elongated objects of size $2\ell$ and zero width (needles), which can rotate within the plane and translate along a straight line.
Each needle is described by a positional $x_i$ and an orientational variable $\phi_i\in [-\pi/2,\pi/2]$.
In the original approach by Kantor and Kardar~\cite{KantorKar2009a}, the needles interact via excluded-volume repulsions only.
In the present work we consider the effects of harmonic elastic interactions as well (represented by blue springs in the figure).

In Sec.~\ref{sec:Model}, we describe our one-dimensional elastic model of hard needles and formulate the statistical problem in the pressure (stress) ensemble.
In Sec.~\ref{sec:gas}, we regain the known results by Kantor and Kardar~\cite{KantorKar2009a} and describe the critical behavior of the nematic order parameter (which was not considered in Ref.~\cite{KantorKar2009a}).
In Sec.~\ref{sec:elastomer}, we obtain a number of properties of this novel one-dimensional nematic elastomer.
Some final considerations and possible extensions of the calculations are given in the conclusions.

\section{The model}
\label{sec:Model}

We consider the model Hamiltonian
\begin{equation}
    \mathcal{H} = \mathcal{H}_{s}+\mathcal{H}_{e},
\end{equation}
where $\mathcal{H}_{s}$ represents the hard-core steric repulsion between needle-like objects,
\begin{eqnarray}
\mathcal{H}_s = \sum_{i=1}^N V_{i-1,i},
\end{eqnarray}
and where $V_{i-1,i}$ is the excluded-volume hard repulsion term,
\begin{eqnarray}
V_{i-1,i} = \left\{
\begin{array}{ll}
0, & \text{ if } x_i - x_{i-1} > \ell \, d_{i-1, i}, \\
\infty, & \text{ otherwise,}
\end{array}
 \right.
 \label{eq:vhn}
\end{eqnarray}
with
\begin{eqnarray}
d_{i-1,i} = \frac{\sin \left\vert \phi_i - \phi_{i-1}\right\vert}{\max \left(\cos \phi_i, \cos \phi_{i-1} \right)}.
\label{eq:dhn}
\end{eqnarray}
The local elastic interactions are given by
\begin{equation}
\mathcal{H}_e = \frac{k}{2} \sum_{i=2}^N \left(x_i - x_{i-1} - a \, \ell \right)^2,
\end{equation}
where we have introduced the elastic energy constant $k>0$, and the equilibrium spacing $a \, \ell$, where $a$ is a dimensionless quantity and $\ell$ is half the length of the needle (see Fig.~\ref{fig:needles}).
Note that our model reduces to a one-dimensional lattice of hard rotors in the limit of infinite $k$.
The statistical properties of orientable objects in one-dimensional lattices have been investigated in diverse contexts, for several forms of hard~\cite{BenmessabihChe2022} and soft-core~\cite{SaryalDha2018} potentials, as well as for other types of anisotropic shapes~\cite{CaseyRun1969}.

\begin{figure}[!ht]
\includegraphics[width=\linewidth]{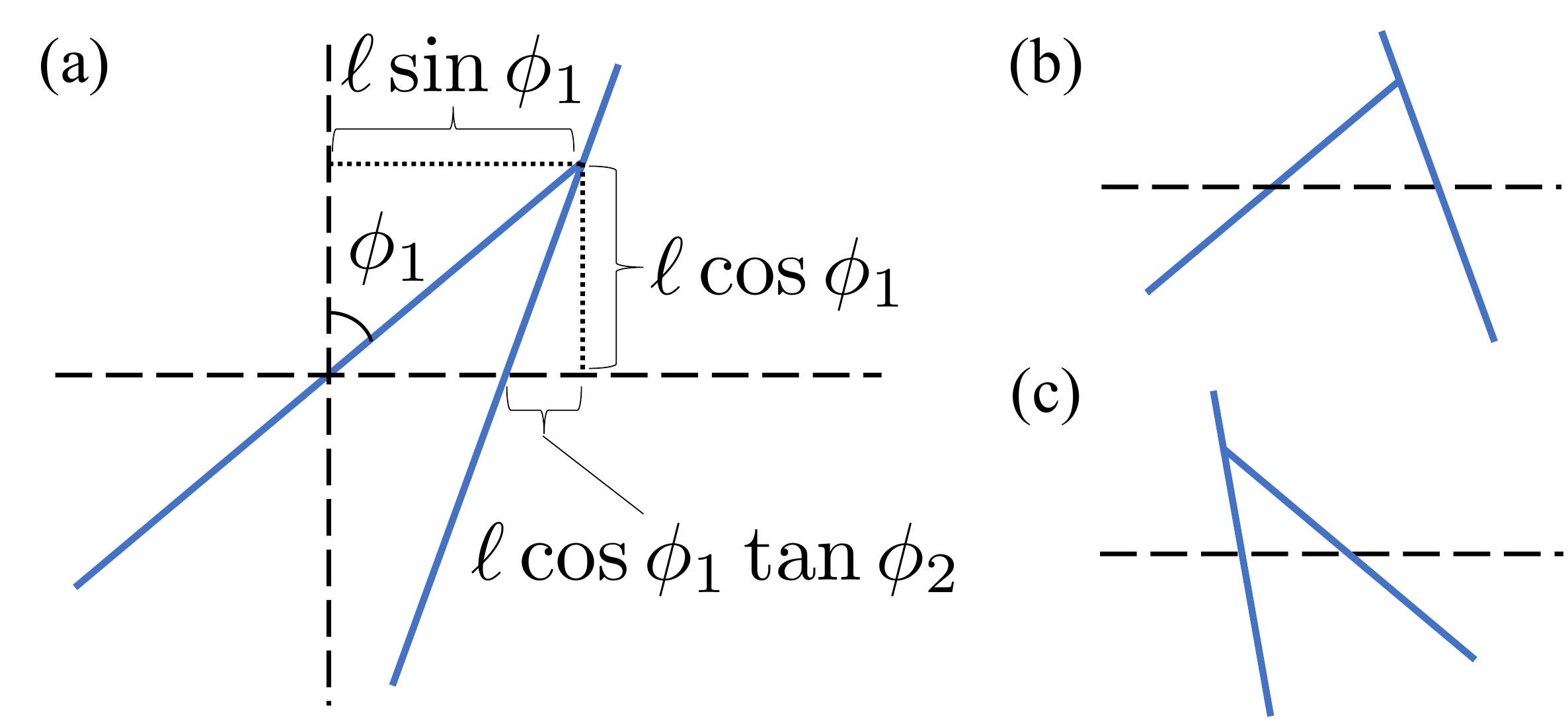}
\caption{Geometry of two adjacent needles at the distance of closest approach for $\phi_1>\phi_2$.
(a), (b) and (c) correspond to the cases for which $\phi_1,\phi_2>0$, $\phi_1>0$ but $\phi_2<0$, and $\phi_1,\phi_2<0$, respectively.
\label{fig:distance}}
\end{figure}

Figure~\ref{fig:distance} shows a derivation of Eq.~\eqref{eq:dhn} for a particular case with $\phi_1>\phi_2$.
When $\phi_1,\phi_2>0$, one can infer from Fig.~\ref{fig:distance}(a) that the distance of closest approach is given by $d_{1,2} / \ell= \sin \phi_1 -\cos \phi_1 \tan \phi_2 = \sin (\phi_1-\phi_2) / \cos \phi_2$.
Similar calculations can be made to show that Eq.~\eqref{eq:dhn} also applies to the cases in which $\phi_1>0$ and $\phi_2<0$ (b), and $\phi_1, \phi_2 <0$ (c), as well as the cases with $\phi_1<\phi_2$.

We now turn to the pressure (stress) ensemble.
In this case, we have $0<x_1<x_2< \cdots < x_{N-1} < x_N \equiv L < \infty$, where $L$ is the system size, which is not fixed.
We also consider free boundary conditions, so that the first and the last particles only interact with particles at their right and left, respectively, and also $V_{0,1} = V_{N,N+1}=0$.
The configurational contribution to the partition function can be written as
\begin{eqnarray}
Y &=& Y \left(\beta, \sigma, N \right) 
\nonumber \\
&=& \int \mathcal{D}x \mathcal{D}\phi \exp \left[ -\beta \left( \mathcal{H} + \sigma \, x_N \right) \right],
\label{eq:z_model}
\end{eqnarray}
where $\beta$ is the inverse temperature, $\mathcal{D} x = \ell^{-N} \prod_{i=1}^N dx_i$ and $\mathcal{D} \phi = \prod_{i=1}^N d\phi_i$ are measures in the positional~\footnote{We used $\ell^{-N}$ in the spatial measure so that $Y$ is dimensionless.
In classical statistical mechanics, it is usual to consider the measure in phase space $d \mu = h^{-N} \prod_i dx_i d p_i$, with the inclusion of Planck's constant $h$, where $p_i$ is the momentum of particle $i$.
This ensures a dimensionless partition function and agreement with the classical limit of an analogous quantum system.
In our case, we could combine $\ell$ with h so that the contribution from the momentum variables is also dimensionless.} and orientational variables.
The last term in the argument of the exponential in \eqref{eq:z_model} is the usual ``pressure'' term, where $\sigma$ is a uniaxial tension.

\section{Hard-needle gas}
\label{sec:gas}

We now recover previous results for a gas of hard needles without elastic interactions (in other words, with the elastic energy parameter $k=0$).
The configurational contribution to the partition function is now given by
\begin{eqnarray}
Y = \int \mathcal{D}x \mathcal{D}\phi \exp \left[ -\beta \left( \sum_{i=1}^N V_{i-1, i} + \sigma \, x_N \right) \right].
\label{eq:gas_model}
\end{eqnarray}
This model has been proposed and solved in Ref.~\cite{KantorKar2009a}.
Interactions can be decoupled with a simple linear transformation,
\begin{eqnarray}
s_i \equiv x_i - x_{i-1},
\end{eqnarray}
so that,
\begin{eqnarray}
\sum_{i=1}^N s_i = x_N = L,
\quad
\frac{\partial \left( x_1, \cdots, x_N \right)}{\partial \left( s_1, \cdots, s_N \right)}=1,
\end{eqnarray}
and the partition function is given by
\begin{eqnarray}
Y &=& \ell^{-N} \int_{0}^{x_2} dx_1 \int_{x_1}^{x_3} dx_2 \cdots \int_{x_{N-1}}^\infty dx_N \int \mathcal{D} \phi
\nonumber \\ && \quad \times \exp \left[-\beta \left( \sum_{i=1}^N V_{i-1,i} + \sigma \, x_N \right) \right]
\nonumber \\ &=& 
\ell^{-N} \int_{0}^{\infty} ds_1 \int_{0}^{\infty} ds_2 \cdots \int_{0}^\infty ds_N \int \mathcal{D} \phi
\nonumber \\ && \quad \times \exp \left[-\beta \sum_{i=1}^N \left( V_{i-1,i} + \sigma \, s_i \right) \right].
\label{eq:zhn1}
\end{eqnarray}
Now we can integrate \eqref{eq:zhn1} over the $s$ variables in order to obtain
\begin{eqnarray}
Y &=& \ell^{-N} \int \mathcal{D} \phi \, \prod_{i=1}^N  \left( \int_{\ell \, d_{i-1,i}}^\infty d s_i \, e^{-\beta \, \sigma \, s_i} \right)
\nonumber \\ &=& (\beta \, \sigma \, \ell )^{-N} \int \mathcal{D} \phi \, \exp \left( - \beta \, \sigma \, \ell \sum_{i=1}^{N} d_{i-1,i} \right).
\end{eqnarray}
Note that our boundary conditions require that $d_{0,1}=0$.

In order to integrate over the angle variables, we consider a discrete set of angles and use the transfer matrix method to evaluate the sum over states. Let us partition the interval $[-\pi/2 , \pi/2]$ into $M$ equal parts, so that the angle variable can be written as
\begin{eqnarray}
\phi^{(k)} = -\frac{\pi}{2} + \frac{k\pi}{M}, \quad \text{with } k=0,1,\cdots, M-1.
\end{eqnarray}
Here we focus on values of $M$ that are large enough to be compatible with the case of continuous orientations; see Refs.~\cite{GurinVar2011,GurinVar2022} for detailed analyses of this model and some variants at small $M$.
We then define a finite-dimensional $M \times M$ transfer matrix $D=((D_{\mu\, \nu}))$, with
\begin{eqnarray}
D_{\mu \, \nu} = \exp \left[ - \tau \, d( \phi^{(\mu)}, \phi^{(\nu)} ) \right],
\end{eqnarray}
where $d$ is the minimum distance between nearest neighbors defined in \eqref{eq:dhn}, and we have introduced the dimensionless tension
\begin{eqnarray}
\tau = \beta \, \sigma \, \ell.
\label{eq:dimfree_pressure}
\end{eqnarray}
Also,
\begin{eqnarray}
\int d \phi_i \approx \sum_{k} 
\left[\phi_i^{(k+1)}-\phi_i^{(k)}\right]
= \frac{\pi}{M} \sum_{k},
\end{eqnarray}
which leads to
\begin{eqnarray}
\prod_{i}\left( \int_{-\frac{\pi}{2}}^{\frac{\pi}{2}}  d \phi_i \, e^{- \tau \, d_{i-1,i}} \right) &\approx& \left(\frac{\pi}{M}\right)^{N} \sum_{k_1=0}^{M-1} \cdots \sum_{k_N=0}^{M-1} D_{k_1\, k_2} 
\nonumber \\ && \quad
\cdot D_{k_2\, k_3} \cdots D_{k_{N-1}\, k_N}
\nonumber \\ &=& \left(\frac{\pi}{M}\right)^N \sum_{k_1, k_N} \left( D^{N-1} \right)_{k_1 \, k_N}. \nonumber
\end{eqnarray}
Thus,
\begin{eqnarray}
Y = \left( \frac{\pi}{M \tau} \right)^N \sum_{\mu, \nu} \left( D^{N-1}\right)_{\mu \, \nu}.
\label{eq:Ygas}
\end{eqnarray}

The free energy density $g = g (\tau )$ is obtained from Eq.~\eqref{eq:Ygas},
\begin{eqnarray}
-\beta g = \ln \left( \frac{\pi}{M \tau} \right) + \frac{1}{N} \ln \left[ \sum_{\mu, \nu} \left( D^{N-1}\right)_{\mu \, \nu}\right].
\end{eqnarray}
This formula becomes simpler in the thermodynamic limit if we consider the similarity transformation
\begin{eqnarray}
B^{-1} \cdot D \cdot B = \text{diag}(\lambda_{1}, \cdots, \lambda_{M}),
\end{eqnarray}
where the components of the $M \times M$ matrix, $B=((B_{\beta \, \alpha}))$, are given by
\begin{eqnarray}
B_{\beta \, \alpha} =  \nu_{\beta \, \alpha},
\end{eqnarray}
where the vector $\boldsymbol{\nu}_{\alpha}\equiv (\nu_{1 \, \alpha}, \cdots, \nu_{M \, \alpha})$ is the $\alpha$-th normalized eigenvector of $D$, with corresponding eigenvalue $\lambda_{\alpha}$. Thus,
\begin{eqnarray}
\sum_{\mu,\nu} \left(D^{N-1}\right)_{\mu \, \nu} = \sum_{\alpha=1}^{M} \left( \sum_{\beta=1}^M \nu_{\beta \,\alpha }\right)^2 {\lambda_{\alpha}}^{N-1}.
\end{eqnarray}
In the thermodynamic limit, the main contribution to this sum comes from the largest eigenvalue $\lambda_{\alpha^*}$, so that the free energy is given by
\begin{eqnarray}
g = -\frac{1}{\beta} \ln \left( \frac{\pi \lambda_{\alpha^*}}{M \tau}\right).
\label{eq:fG_HN}
\end{eqnarray}

We now remark that one-point averages of a quantity $X_{k_i}$ can be directly calculated using the definition
\begin{eqnarray}
\langle X_{k_i} \rangle &=& \frac{\sum_{k_1, \cdots, k_N } X_{k_i} D_{k_1 \, k_2} \cdots D_{k_{N-1} \, k_N}}{\sum_{k_1, \cdots, k_N} D_{k_1 \, k_2} \cdots D_{k_{N-1} \, k_N}}
\nonumber \\ &=&
\frac{\sum_{\mu,\nu} \left(D^{i-1} \cdot X \cdot D^{N-i} \right)_{\mu \, \nu}}{\sum_{\mu,\nu} \left( D^{N-1} \right)_{\mu \, \nu}},
\label{eq:Xaverage}
\end{eqnarray}
where $X$ in the second line denotes a one-point $M \times M$ matrix,
\begin{eqnarray}
X=((X_{\mu \nu})) = \text{diag}(X_1, \cdots, X_M).
\end{eqnarray}
Using the basis of eigenvectors of $D$, Eq.~\eqref{eq:Xaverage} can be rewritten in the thermodynamic limit as~\footnote{Note that Eq.~\eqref{eq:xk_HN} is valid for bulk particles.
For boundary particles, $< X_{k_i} > =
(\sum_{\beta=1}^M \nu_{\beta \, \alpha^*}X_{\beta})/(\sum_{\beta=1}^M \nu_{\beta \, \alpha^*})$.}
\begin{eqnarray}
\langle X_{k_i} \rangle = \sum_{\beta=1}^M {\nu_{\beta \, \alpha^*} }^2 X_{\beta}. 
\label{eq:xk_HN}
\end{eqnarray}
For planar orientations, the nematic order parameter $Q$ is given by
\begin{eqnarray}
Q = \langle \cos \left( 2 \, \phi_i \right) \rangle,
\label{eq:nematic_order}
\end{eqnarray}
which can be numerically evaluated by means of Eq. \eqref{eq:xk_HN}.

We now turn to the average distance between needles, which can be written as
\begin{align}
s & \equiv \frac{1}{N} \, Y^{-1} \int \mathcal{D}x  \, \mathcal{D}\phi \, x_N \, e^{-\beta \mathcal{H} - \beta \sigma x_N} \nonumber \\
& = - \frac{1}{N \beta} Y^{-1} \frac{\partial Y}{ \partial \sigma }
= \ell \, \frac{\partial (\beta g)}{ \partial \tau}.
\end{align}

\begin{figure}[!ht]
\centering
\includegraphics[width=\linewidth]{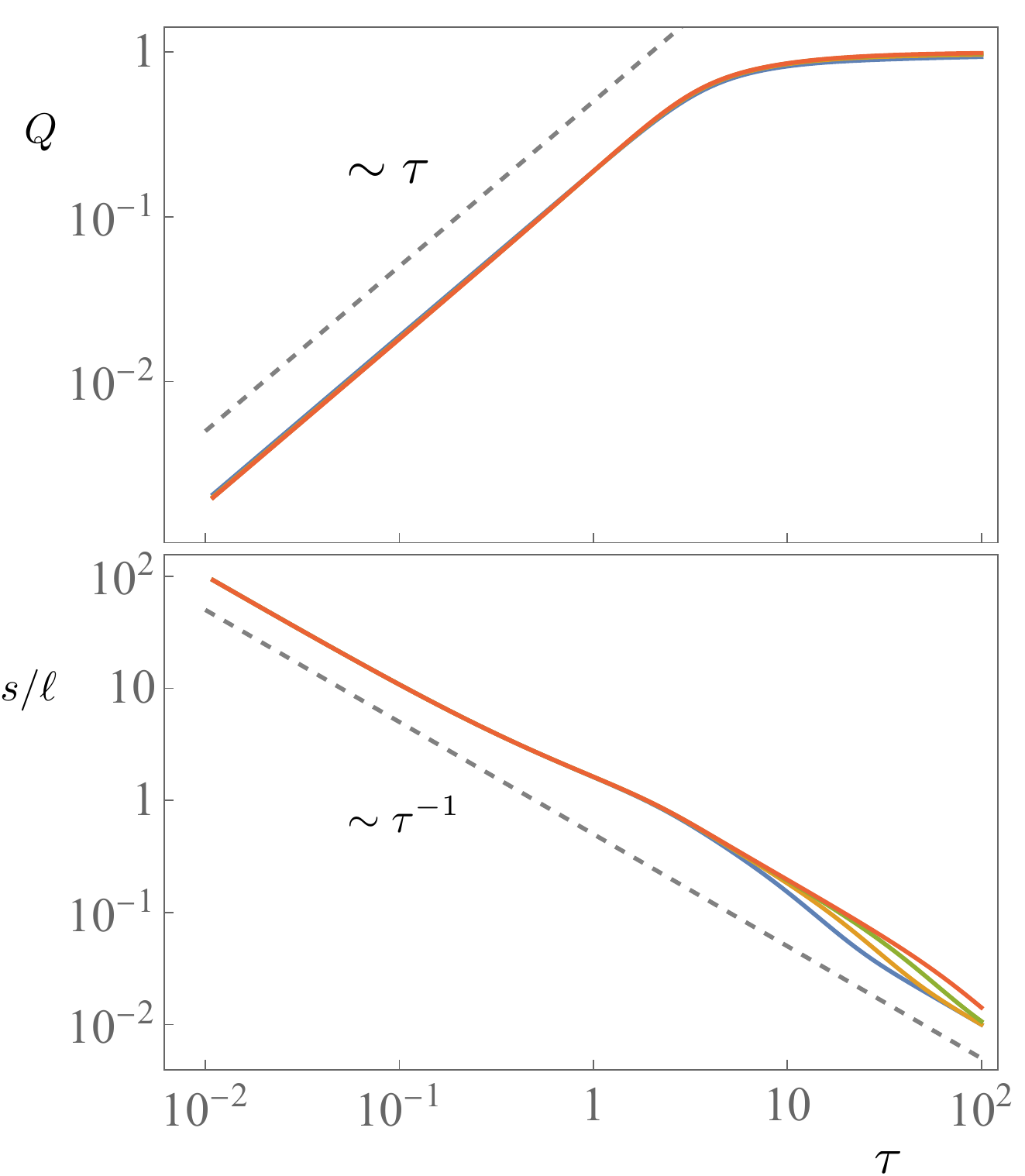}
\caption{Nematic order parameter $Q$ and average spacing between needles $s/\ell$ as a function of the dimensionless tension $\tau=\beta \,\sigma \, \ell$ for a gas of hard needles without elastic interactions (i.e. with $k=0$).
Blue, yellow, green and red lines correspond to $M=16$, $32$, $64$ and $128$, respectively.
The dashed lines on the top and bottom plots correspond to power laws with exponent $1$ and $-1$, respectively.
\label{fig:qsxtau}}
\end{figure}

In Figure~\ref{fig:qsxtau} we show the nematic order parameter $Q$ and the average spacing between needles $s$ as a function of the dimensionless tension $\tau=\beta \, \sigma \, \ell$ for $M=16$ (blue), $32$ (yellow), $64$ (green) and $128$ (red).
Notice that the discrete angle approximation quickly converges even for modest values of $M$.
As it should be anticipated, there is no long-range order in the absence of tension.
As indicated by gray dashed lines, the order parameter vanishes as $\tau$, and the average spacing between needles diverges as $\tau^{-1}$ in the limit of small $\tau$.
We remark that our results agree with both the previous publication by Kantor and Kardar~\cite{KantorKar2009a} and with some unpublished numerical simulations performed by Annunziata and Petri.

\section{Hard-needle elastomer}
\label{sec:elastomer}

We now consider the elastic case, with $k\neq 0$.
Using our previous change of variables $x_i \rightarrow s_i=x_i - x_{i-1}$, we can write the partition function in the pressure ensemble as
\begin{eqnarray}
Y &=& Y (\beta, \sigma, N)
\nonumber \\
&=& \ell^{-N} \int \mathcal{D} \phi \prod_{i=1}^N \left\{ \int_{\ell \, d_{i-1,i}}^\infty d s \exp \left[ -\frac{\beta k}{2} (\delta_{i,1} -1)
\right. \right. \nonumber \\ && \quad \left. \left.
 \times (s - a \, \ell )^2 
- \beta \, \sigma \, s \right] \right\}, \nonumber
\end{eqnarray}
where the Kronecker delta in the exponent ensures the free boundary condition, i.e. there is only one spring attached to the first particle.
It is convenient to introduce another dimensionless parameter,
\begin{eqnarray}
\Lambda \equiv \beta \, k \, {\ell}^2,
\end{eqnarray}
so that, with some algebra, we have
\begin{eqnarray}
Y &=& \exp \left\{ N \left[ \frac{1}{2}\ln \left( \frac{\pi }{2\Lambda}\right) + \tau \left(\frac{\tau}{2\Lambda} -a \right)\right] \right\}  \int \mathcal{D} \phi 
\nonumber \\ && \quad \times \prod_{i=2}^N \left\{ \text{erfc} \left[ \sqrt{\frac{\Lambda}{2}} \left(d_{i-1,i} -a + \frac{\tau}{\Lambda}\right) \right] \right\},
\end{eqnarray}
where $\tau$ has been defined in Eq. \eqref{eq:dimfree_pressure}, erfc$(x)$ is the complementary error function, and we have neglected terms of order $\mathcal{O}(1)$. The transfer matrix is now given by
\begin{eqnarray}
D_{\mu \, \nu} = \text{erfc} \left\{ \sqrt{ \frac{\Lambda}{2}} \left[ d\left( \phi^{(\mu)}, \phi^{(\nu)} \right) -a + \frac{\tau}{\Lambda} \right] \right\},
\label{eq:TMelastomer}
\end{eqnarray}
and the free energy, $g (\beta, \sigma) = g(\tau)$, can be obtained from
\begin{eqnarray}
-\beta g = \frac{1}{2} \ln \left[ \frac{\pi^3}{2M\Lambda} ( {\lambda_{\alpha^*}})^2 \right] + \tau \left(\frac{\tau}{2\Lambda} - a \right),
\end{eqnarray}
where $ {\lambda_{\alpha^*}}$ is the largest eigenvalue of $D$. We can use Eqs.~\eqref{eq:xk_HN} and the eigenvectors of $D$, given by Eq.~\eqref{eq:TMelastomer}, to calculate one-point averages.

In the previous section, we have shown that the dominant singularity at $\tau = 0$ leads to power-law behavior of the nematic order parameter ($Q \sim \tau$) and of the average spacing between needles ($s \sim \tau^{-1}$). 
For nonzero $k$ (or $\Lambda$), invariant scaling behavior suggests that $Q$ and $s$ are not functions of $\tau$ and $\Lambda$ independently, and we anticipate that
\begin{equation}
Q = \vert\tau \vert \, \mathcal{Q} (\Lambda \,\vert\tau\vert^{-\Delta})
\label{eq:ElastomerQscaling}
\end{equation}
and similarly~\footnote{The average spacing $s$ can be calculated from a derivative of the free energy as $s/\ell = \partial (\beta f) / \partial \,\tau$.
For the hard-needle elastomer, this calculation results in $s/\ell = -(1/\lambda_{\alpha^*}) \partial \lambda_{\alpha^*} / \partial \, \tau + a - \tau / \Lambda$.
The first term yields the expected scaling behavior, whereas the last term provides important corrections when $\Lambda \sim \tau^2$.} that $s = \vert\tau\vert^{-1} \mathcal{S} (\Lambda / \vert\tau\vert^\Delta)$ where $\mathcal{Q}$ and $\mathcal{S}$ are universal scaling functions~\cite{SethnaZap2017}, and $\Delta$ is a critical exponent.
Although critical exponents have been historically considered the paradigm of universal behavior, we emphasize that many other quantities are universal besides the exponents~\cite{Cardy1996}.
Notorious examples include amplitude ratios, which often provide a better test of universality classes than do critical exponents~\cite{ChaikinLub1995}.
An interesting and open question is the determination of what features of a ``universal'' scaling function such as $\mathcal{Q}$ are indeed universal.

\begin{figure}[!ht]
\centering
\includegraphics[width=\linewidth]{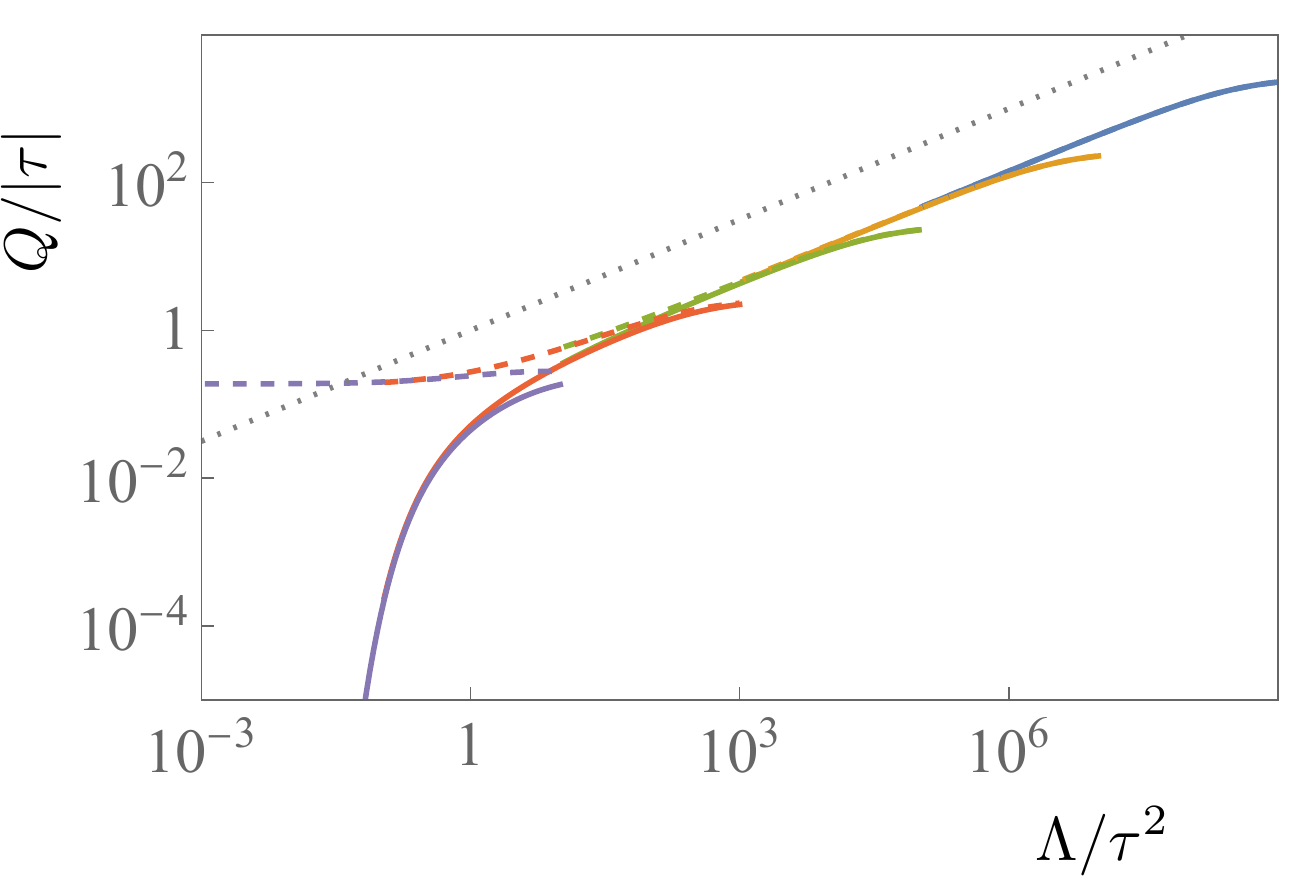}
\caption{Scaling collapse plots for the rescaled nematic order parameter as a function of rescaled $\Lambda=\beta \, k \, \ell^2$ for a hard-needle elastomer with $a=1$, several values of $\tau$ (different colors) and for both compression $\tau>0$ (dashed curves) and dilation $\tau < 0$ (solid curves).
The gray-dotted line was inserted to indicate power-law behavior $\sim x^{1/2}$ at large $\Lambda / \tau^2$.
\label{fig:qxLambda_Scaled}}
\end{figure}

To validate the universal scaling form encapsulated by Eq.~\eqref{eq:ElastomerQscaling}, Figure~\ref{fig:qxLambda_Scaled} shows scaling collapse plots for the rescaled nematic order parameter $Q / \vert\tau\vert$ as a function of $\Lambda / \vert\tau\vert^\Delta$, for $M=128$, $a=1$ and $\vert\tau\vert=10^{-4}$ (blue), $10^{-3}$ (yellow), $10^{-2}$ (green), $10^{-1}$ (red), and $1$ (purple).
Here we consider both positive (compression) and negative (dilation) tension, corresponding to dashed and solid lines, respectively.
We have varied the critical exponent $\Delta$ until we find that the curves collapse into two branches (corresponding to $\tau > 0$ and $\tau < 0$) of a single universal curve when $\Delta = 2$.
Different values of $a$ do not affect the overall scaling behavior.

For very stiff systems (i.e. for large $k$), one expects the overall behavior to be dominated by the value of the spring constant, and to show only a small dependence on the stress.
This physical intuition is corroborated by our results; at large $\Lambda / \tau^2$, Fig.~\ref{fig:qxLambda_Scaled} indicates that the universal function $\mathcal{Q}(x) \sim x^{1/2}$, so that $Q\sim \sqrt{\Lambda}$ independent of $\tau$.
It is worth noting that $\Lambda \approx 1$ marks a threshold above which corrections to scaling become important (indicated by the curves peeling off from the putative universal function at large arguments).
At small values of $\Lambda / \tau^2$, compression and dilatation lead to very different outcomes.
Whereas $Q \sim \tau$, independent of $\Lambda$, for compression at low $\Lambda$, the system exhibits a strikingly sharp decay for dilatation at low $\Lambda$. 

We now turn to the system scaling behavior at the critical value $\tau=0$.
An alternative expression for the scaling form given by Eq.~\eqref{eq:ElastomerQscaling} can be obtained using a simple change of variables,
\begin{align}
Q & = \vert\tau \vert \, \Lambda^{-1/\Delta} \Lambda^{1/\Delta} \mathcal{Q} (\Lambda \,\vert\tau\vert^{-\Delta}) \nonumber \\
& = \Lambda^{1/\Delta} \tilde{\mathcal{Q}} (\vert\tau\vert \, \Lambda^{-1/\Delta}),
\label{eq:ElastomerQscaling2}
\end{align}
where $\tilde{\mathcal{Q}}(x)= x\,\mathcal{Q}(x^{-\Delta})$ is a new scaling function.
Note that Eq.~\eqref{eq:ElastomerQscaling2} implies that $Q \sim \Lambda^{1/\Delta}$ as $\tau \rightarrow 0$.
\begin{figure}[!ht]
\centering
\includegraphics[width=\linewidth]{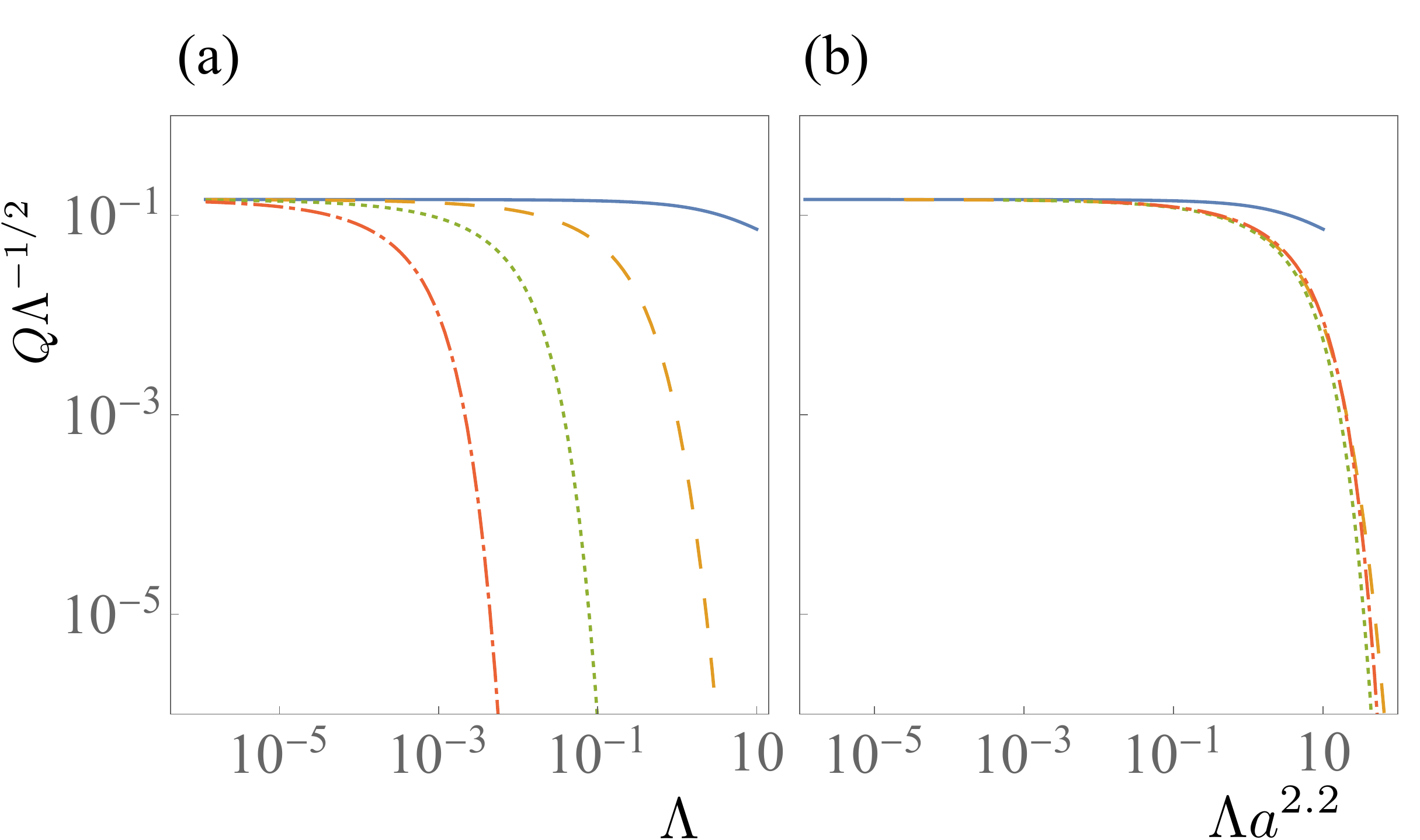}
\caption{(a) Rescaled $Q$ as a function of $\Lambda$ at $\tau = 0$ for several values of $a$ (different colors).
(b) The same as $a$, but with rescaled $\Lambda a^{2.2}$ in the x-axis.
\label{fig:qxLambdaZerotau_Scaling}}
\end{figure}

Figure~\ref{fig:qxLambdaZerotau_Scaling}(a) shows a plot of the rescaled order parameter $Q \Lambda^{-1/2}$ as a function of $\Lambda$ for $M=128$ and $a = 1$ (blue, solid), $4$ (yellow, dashed), $16$ (green, dotted) and $64$ (red, dash-dotted).
Notice the steep crossover to a solution with $Q \approx 0$ at larger values of $\Lambda$.
Also notice that all curves have approximately the same shape for large $a$, which suggests that a scaling combination of $\Lambda$ and $a$ may collapse these curves.
We then show in Fig.~\ref{fig:qxLambdaZerotau_Scaling}(b) a plot of rescaled $Q$ as a function of rescaled $\Lambda a^\rho$, with $\rho = 2.2$ chosen so that we obtain the best scaling collapse for large $a$ data.
Since $a$ is ratio of two characteristic length scales of the system, it is not surprising that invariant scaling combinations involving $a$ are present in some regimes.

\section{Conclusions}
\label{sec:conclusions}
We have obtained a number of analytic expressions and numerical results describing the statistical behavior of a one-dimensional system of hard needles with the inclusion of steric repulsions and elastic interactions.
Using the transfer matrix technique, we have exactly calculated the partition function, free energy and order parameter for this model.
We have then discussed the system critical scaling behavior, and described a standard universal scaling form that is controlled by a putative zero-tension fixed point.
The rescaled order parameter $Q / \vert\sigma\vert$ can be written as a universal function of rescaled elastic energy constant $k / \vert\sigma\vert^\Delta$, with $\Delta$ denoting a critical exponent that is equal to $2$ for this model.

In future work, we plan to consider other forms of interacting potentials, including competing terms~\cite{BienzobasSan17} or chiral twist terms~\cite{NascimentoSal2019}, which are known to lead to modulated phases and that would allow us to make contact with the nematic cholesteric behavior~\cite{Gennes1995}.
We also plan to use the coherent potential approximation~\cite{FengGar1985,LiarteLub2019} to incorporate disorder in the elastic variables, which would also be relevant in the context of jamming~\cite{LiuNag2010} and other classes of rigidity transitions~\cite{LiarteSet2022}.
Finally, it would be interesting to investigate the interplay between elasticity and excluded volume steric interactions in generalizations of the models considered in Refs.~\cite{SaryalDha2022,KlamserDha2022}.

\begin{acknowledgments}
DBL thanks the financial support through FAPESP grants \verb|#| 2021/14285-3 and \verb|#| 2022/09615-7.
AP is grateful to ICTP-SAIFR for a two-months visiting grant.
\end{acknowledgments}

%

\end{document}